# Anisotropy of Arrival Directions of $E_0 \geqslant 8 \times 10^{18}$ eV Cosmic Rays and Cosmic Microwave Background

A. V. Glushkov*

*Institute of Cosmophysical Research and Aeronomy, Yakutsk Research Center, Siberian Branch, Russian Academy of Sciences, pr. Lenina 31, Yakutsk, 677980 Russia*

**Abstract**—Results are presented that were obtained by analyzing the arrival directions of $E_0 \geqslant 8 \times 10^{18}$ eV primary cosmic rays recorded at the Yakutsk array over the period between 1974 and 2003 and at the SUGAR array (Australia). The greatest primary-cosmic-ray flux is shown to arrive from the region of visible intersection of the planes of the Galaxy and the Supergalaxy (local supercluster of galaxies) at a galactic longitude of about 137°. On a global scale, the lowest temperature of the cosmic microwave background is typical of this region.

## 1. INTRODUCTION

Giant air showers (GAS) of limiting energy $E_0 \geqslant 10^{19}$ eV have attracted great interest since the first events recorded at the largest world facilities, including the Volcano Ranch (USA) [1], the Haverah Park (UK) [2], SUGAR (Australia) [3] and the Yakutsk arrays [4]. This interest is caused largely by a sharp variation observed at all of these facilities in the behavior of the spectrum of primary cosmic rays in this energy region. Many experimental and theoretical studies have addressed the problem of the GAS origin, but it still remains one of the most intricate problems.

The anisotropy of primary cosmic rays contains important information about the GAS origin. These events are indicative of a slight correlation between their arrival directions and the Galaxy plane (see [5–7], for example) and a more pronounced correlation between the arrival directions in question and the plane of the Supergalaxy (local supercluster of galaxies) [8–2]. Observations of clusters in the GAS arrival directions were reported in [12–14]. A correlation between some of the GASs and galaxies having an active core was found in [14–16]. According to [16–19], GASs may be produced by neutral primary particles. In the following, we will consider some features that are peculiar to the GAS anisotropy and to the cosmic microwave background and which can shed more light on the problem of the origin of ultrahigh-energy primary cosmic rays.

*E-mail: a.v.glushkov@ikfia.ysn.ru

## 2. RESULTS OF OBSERVATIONS

Figure 1 shows the GAS density distribution on a map of the developed celestial sphere in terms of galactic coordinates. In order to construct it, we used $E_0 \geqslant 8 \times 10^{18}$ eV events recorded by the Yakutsk array between 1974 and 2003 at zenith angles in the range $\theta \leqslant 60°$. For our analysis, we selected only those events for which the arrival directions were fixed on the basis of data from more than four stations and whose axes were within the array perimeter. This enabled us to measure the GAS arrival directions to a precision not poorer than 3°. The primary-particle energy $E_0$ was obtained from the relations

$$E_0 = (4.8 \pm 1.6) \times 10^{17} \quad (1)$$
$$\times (\rho_{s,600}(0°))^{1.0 \pm 0.02} \text{ [eV]},$$
$$\rho_{s,600}(0°) = \rho_{s,600}(\theta) \quad (2)$$
$$\times \exp(1020(\sec\theta - 1)/\lambda_\rho) \text{ [m}^{-2}\text{]},$$
$$\lambda_\rho = (450 \pm 44) + (32 \pm 15) \quad (3)$$
$$\times \log(\rho_{s,600}(0°)) \text{ [g/cm}^2\text{]},$$

where $\rho_{s,600}(\theta)$ is the charged-particle density measured with ground-based scintillation detectors at a distance of $R = 600$ m from the shower axis to precision of about 30%. In all, we selected 519 showers in this manner.

In addition, we used 522 $E_0 \geqslant 8 \times 10^{18}$ eV showers from the catalog presented in [20], which were recorded by the SUGAR facility. Their angular precision is not poorer than 5°. These showers refer mostly to the southern hemisphere of the Earth. Together with the data from the Yakutsk array, they provide quite a comprehensive pattern of the anisotropy of ultrahigh-energy primary cosmic rays in the surrounding space. Of course, data from other world facilities would increase GAS statistics, but, unfortunately, they are mostly absent from free-access catalogs, this rendering them unavailable to other researchers for analysis. Therefore, we restrict ourselves to the data obtained at the two above facilities. They provide quite an acceptable uniformity of the GAS arrival directions over the entire sphere for the global anisotropy to be analyzed by the method outlined below.

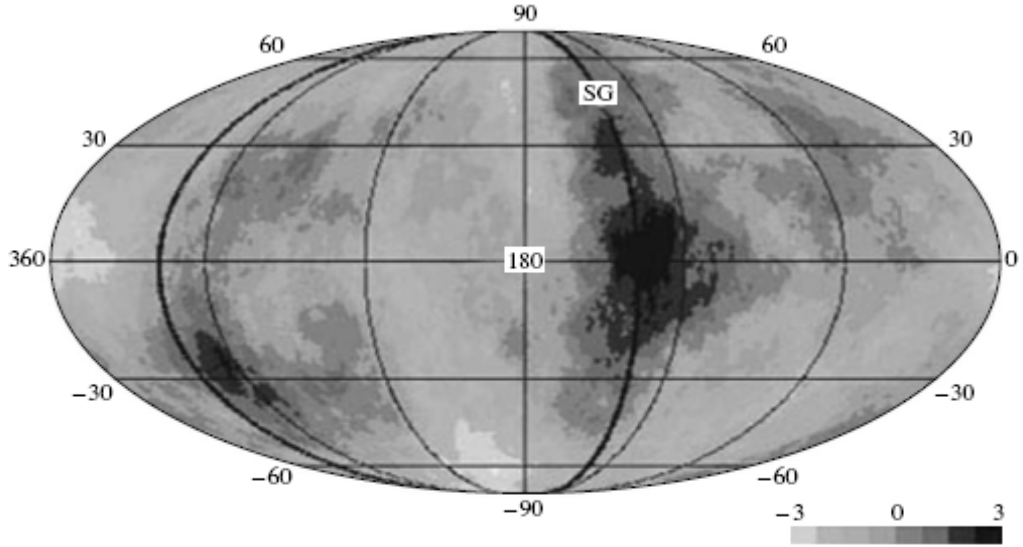

**Fig. 1.** Deviation of the observed number of showers, $N_1$, from the expected average number, $\langle N \rangle$, in units of $n_\sigma = (N_1 - \langle N \rangle)/\sqrt{\langle N \rangle}$. These are plotted on a map of the developed celestial sphere in terms of galactic coordinates for GAS of energy in the region $E_0 \geqslant 8 \times 10^{18}$ eV. The SG curve is the Supergalaxy plane.

We studied deviations of the observed number of events, $N_1$, from the expected average number $\langle N \rangle = N_2(\Omega_1/\Omega_2)$ in units of the standard deviation $\sigma = \sqrt{\langle N \rangle}$; that is,

$$n_\sigma = (N_1 - \langle N \rangle)/\sigma, \qquad (4)$$

where $N_1$ and $N_2$ are the numbers of showers in the solid angles $\Omega_1 = 2\pi(1 - \cos\theta_1)$ and $\Omega_2 = 2\pi(1 - \cos\theta_2)$ ($\theta_1 = 20°$ and $\theta_2 = 60°$), respectively. The values of the deviation in (4) were calculated by successively moving the center of a ($1° \times 1°$) area over the entire sphere. The limits of the measurement of $n_\sigma$ are presented at the bottom of Fig. 1 in the form of a toned scale. The most darkly shaded regions correspond to the GAS flux exceeding the average by $n_\sigma \geqslant 3\sigma$. The Supergalaxy plane is represented by the SG curve.

Figure 1 demonstrates that no excess radiation comes from the Galaxy center, where the most vigorous and intense processes of matter transformation occur. Moreover, there is a relative scarcity of events here. But an excess GAS flux is clearly visible in the region of intersection of the Galaxy and Supergalaxy planes at a galactic longitude of about 137°. It was shown in [21] by the wavelet-analysis method that this local sky region manifests itself most vividly in the arrival directions of $E_0 \approx (1-2) \times 10^{19}$ eV primary cosmic rays (the probability of a random event is 0.007). The pole characterized by a maximum number of events has equatorial coordinates of $\alpha_{max} \approx 35° \pm 20°$ and $\delta_{max} \approx 52.5° \pm 7.5°$ and lies in the Supergalaxy plane. In our opinion, a slight extendedness of this region along the Supergalaxy plane suggests that primary cosmic rays of the above energy are of an extragalactic origin. It was shown in [16–18] that quasars generating neutral particles may be sources of such primary cosmic rays. These particles traverse the Supergalaxy and the Galaxy on their path to the Earth. Probably, some of them are involved in nuclear reactions with the gas of these structures. The highest gas concentration is observed along the visual ray just in the region of visible intersection of the Supergalaxy and Galaxy disks at a galactic longitude of about 137°.

Let us consider yet another intriguing fact that may be directly related to the above result. Figure 2 shows the distribution of the cosmic microwave background over the map of the celestial sphere in terms of galactic coordinates according to the data from the WMAP (Wilkinson Microwave Anisotropy

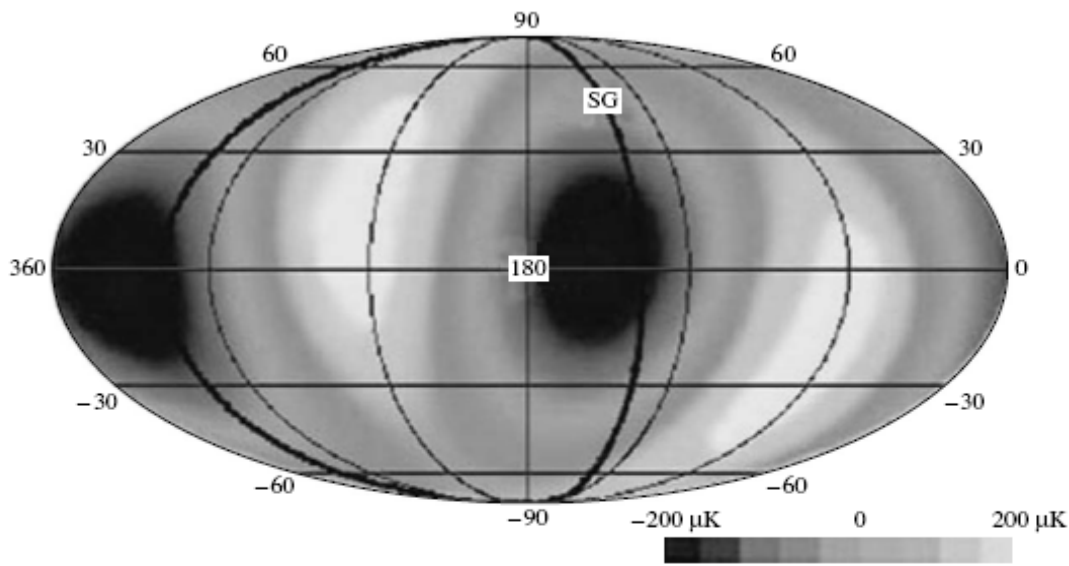

**Fig. 2.** Distribution of the cosmic microwave background over the map of the celestial sphere in terms of galactic coordinates according to the data from the WMAP aircraft for the quadrupole moment $l = 2$ of the Fourier expansion of the background temperature in spherical harmonics [22]. The most darkly and most lightly shaded regions correspond to deviations of, respectively, $-200$ and $+200$ $\mu$K from the average temperature of 2.725 K. The SG curve is the Supergalaxy plane.

Probe) aircraft. The quadrupole moment ($l = 2$) obtained from the Fourier expansion of the background temperature in spherical harmonics [22] is presented there. The main harmonic ($l = 0$), a monopole, characterizes the average relic temperature of 2.725 K over the entire sphere. The next harmonic ($l = 1$) is a dipole, in which the temperature is higher in one of the hemispheres and is lower in the other. The dipole arises because of the Doppler effect since the Solar System moves with respect to the relic—the sky is warmer in the direction of the Sun's motion. In Fig. 2, the most darkly and most lightly shaded regions correspond to deviations of $-200$ and $+200$ $\mu$K from the average temperature. The SG curve represents the Supergalaxy plane.

It is evident that the SG curve intersects the Galaxy plane approximately in the coldest regions of the microwave background. This fact cannot be explained within the standard inflationary theory of Universe formation via the Big Bang. We will not dwell on this or some other contradictions of fundamental importance between the theory and experiment. They were considered in detail elsewhere [22, 23]. We would only like to draw attention to the fact that one of the cold poles in Fig. 2 (at a galactic longitude of about 137°) coincides with the maximum of the flux of ultrahigh-energy cosmic rays in Fig. 1.

## 3. CONCLUSIONS

At the present time, it is difficult to explain the above correlation within the current concepts of the nature of primary particles producing GASs. It cannot be precluded that the above and other new data may change substantially our ideas of the early evolution of the Universe. Some indications of this have already been obtained. For example, the results obtained by studying the anisotropy of various objects characterized by redshifts of $z \leqslant 6$ are presented in [16, 24]. A global anisotropy is also found at the site of these objects; it is caused by a shift of the observer's position relative to the center of their spherical symmetry approximately toward the region of visible intersection of the Galaxy and Supergalaxy discs at a galactic longitude of about 137°. It is possible that the global anisotropy of GAS arrival directions and that of the cosmic microwave background are of the same origin associated with the gravitational potential of the Metagalaxy in which we live. We deem that all of the above facts are of great importance and therefore deserve further study.

REFERENCES
1. J. Linsley, Phys. Rev. Lett. **10**, 146 (1963).
2. D. M. Edge, A. C. Evans, H. J. Garmston, et al., J. Phys. A **6**, 1612 (1973).
3. C. J. Bell et al., J. Phys. A **7**, 990 (1974).

4. D. D. Krasilnikov, A. I. Kuzmin, J. Linsley, et al., J. Phys. A **7**, 176 (1974).
5. J. Szabelski et al., J. Phys. G., **12**, 1433 (1986).
6. B. N. Afanasiev et al., in *Proccedings of the 24th ICRC, Rome, 1995*, Vol. 2, p. 756.
7. A. A. Mikhaĭlov, Pis'ma Zh. Éksp. Teor. Fiz. **72**, 233 (2000) [JETP Lett. **72**, 160 (2000)].
8. T. Stanev, P. L. Bierman, J. Lloyd-Evans, et al., Phys. Rev. Lett. **75**, 3056 (1995).
9. A. V. Glushkov, Pis'ma Zh. Éksp. Teor. Fiz. **73**, 355 (2001) [JETP Lett. **73**, 313 (2001)].
10. A. V. Glushkov, Pis'ma Astron. Zh. **28**, 341 (2002) [Astron. Lett. **28**, 296 (2002)].
11. A. V. Glushkov, Pis'ma Astron. Zh. **29**, 172 (2003) [Astron. Lett. **29**, 142 (2003)].
12. A. V. Glushkov, Izv. Akad. Nauk, Ser. Fiz. **69**, 366 (2005).
13. M. Takeda, N. Hayashida, K. Honda, et al., Astrophys. J. **522**, 225 (1999).
14. P. G. Tinyakov, and I. I. Tkachov, Pis'ma Zh. Éksp. Teor. Fiz. **74**, 499 (2001) [JETP Lett. **74**, 445 (2001)].
15. A. V. Urynson, Zh. Éksp. Teor. Fiz. **116**, 1121 (1999) [JETP **89**, 597 (1999)].
16. A. V. Glushkov, Yad. Fiz. **68**, 262 (2005) [Phys. At. Nucl. **68**, 237 (2005)].
17. A. V. Glushkov and M. I. Pravdin, Zh. Éksp. Teor. Fiz. **130**, 963 (2006) [JETP **103**, 831 (2006)].
18. A. V. Glushkov, Yad. Fiz. **70**, 353 (2007) [Phys. At. Nucl. **70**, 326 (2007)].
19. A. V. Glushkov, D. S. Gorbunov, I. T. Makarov, et al., Pis'ma Zh. Éksp. Teor. Fiz. **85**, 163 (2007) [JETP Lett. **85**, 131 (2007)].
20. M. M. Winn et al., *Catalogue of the Highest Energy Cosmic Rays. Giant Extensive Air Showers, No. 2* (World Data Center C2 for Cosmic Rays, Japan, 1986).
21. A. A. Ivanov, A. D. Krasilnikov, M. I. Pravdin, in *Proccedings of the 28th ICRC, Tsukuba, 2003*, Vol. 1, p. 341.
22. G. Hinshaw, M. R. Nolta, C. L. Bennett, et al., Astrophys. J. (in press).
23. G. Starkman and D. Shvarts, V Mire Nauki **11**, 28 (2005).
24. A. V. Glushkov, astro-ph/0512276.